\documentclass[%
superscriptaddress,
preprint,
 amsmath,amssymb,
 aps,
]{revtex4-1}

\usepackage{graphicx}% Include figure files
\usepackage{dcolumn}% Align table columns on decimal point
\usepackage{bm}% bold math
\begin{document}

\preprint{APS/123-QED}

\title{Expected yields of $^{181}$Ta (e,e') $^{181}$Ta$^*$ in the multi-keV range with a plasma-cathode electron beam}% Force line breaks with \\

\author{F.Gobet}
 \email{gobet@cenbg.in2p3.fr}

\affiliation{Universit\'e Bordeaux, CNRS-IN2P3, CENBG, F-33175 Gradignan, France}%

\author{A.Ya.Dzyublik}

\affiliation{Institute for Nuclear Research, National Academy of Sciences of Ukraine, Kiev, Ukraine}%

\author{G.Gosselin}%

\affiliation{ 
CEA, DIF, F-91297 Arpajon, France
}%

\author{V.M\'eot}%
\affiliation{ 
CEA, DIF, F-91297 Arpajon, France
}%
\affiliation{ 
Universit\'e de Paris-Saclay, CEA, LMCE F-91680 Bruy\`eres-le-ch\^atel, France
}%

 \author{M.Versteegen}
 \affiliation{Universit\'e Bordeaux, CNRS-IN2P3, CENBG, F-33175 Gradignan, France}
 
\date{\today}% It is always \today, today,
             %  but any date may be explicitly specified

\begin{abstract}
Calculations of the cross sections of inelastic electron scattering $(e,e')$ on a nucleus in the multi-keV energy range strongly depend on the description of the screening of the nuclear Coulomb potential as well as on the deformation of the wave functions of the incoming and outgoing electron in the vicinity of the nucleus. These cross sections are evaluated at values lower than $10^{-30}$ cm$^2$ which vary by several orders of magnitude according to the models. Experimental measurements would be required to constrain the models but it is a real challenge to measure such low cross sections.
In this study, we demonstrate that inelastic electron scattering is the main nuclear excitation mechanism in a $^{181}$Ta target irradiated with a new intense 10 - 30 keV electron beam produced with a biased laser-plasma. Calculations show that through the detection of conversion electrons, it should be possible to measure the nuclear excitation yields. The effect of electron beam heating and of plasma deposition on the tantalum target are quantified, thus allowing the dimensioning of a possible experimental configuration to study $(e,e')$ processes in this range of energy for the first time.
\end{abstract}

%Use showkeys class option if keyword
                              %display desired
\maketitle

\section{Introduction}

Dense and hot plasmas constitute 99.9\% of the matter in the Universe, especially in the stellar environment \cite{nishikawa2013plasma,burbidge1957synthesis}. It is in these plasmas that nucleosynthesis processes take place, whereby nuclei heavier than iron are produced by a succession of neutron captures and $\beta$ decays, followed by fission when the nuclei become too heavy \cite{iliadis2015nuclear}. In astrophysical plasma environments, the density of free particles can be high and temperatures can reach values not experienced on Earth \cite{klay1991nuclear}. These extreme conditions favor the transitions to the nuclear excited states via various electromagnetic processes \cite{ichimaru1993nuclear,brown1997nuclear,harston1999mechanisms}.  As neutron capture cross sections depend on the nuclear states \cite{belier2015integral}, knowledge of the excited state distributions in plasmas is required to predict the effects of the thermodynamic conditions on the nucleosynthesis pathways \cite{klay1991nuclear,doolen1978nuclear,gosselin2007modified}.  

Among nuclear excitation processes occurring in astrophysical plasmas, inelastic electron scattering $(e,e')$ on a nucleus plays a particular role. In a plasma, all free electrons with energies higher than the nuclear transition energy can excite the nucleus. The high number of electrons in plasmas can lead to significant nuclear excitation rates, even though the expected cross sections are small, often well under 10$^{-30}$ cm$^2$. The process of electron inelastic scattering has been experimentally and theoretically studied in the MeV energy range\cite{barber1962inelastic,theissen1972spectroscopy,tkalya1991theoretical,arutyunyan1988inelastic} but in the keV range of interest as in astrophysical plasmas\cite{klay1991nuclear} only a few models have been developed to calculate the expected cross sections\cite{letokhov1994excitation,gosselin2009nuclear,dzyublik2013role,tkalya2012cross}. The main theoretical problem lies in the determination of the radial matrix element that describes the incident electron. Its calculation is complex at low energy as it depends on the nuclear Coulomb potential which can be partially screened by surrounding electrons. 

Experimental measurements are therefore necessary to constrain these models. 
The $^{181}$Ta nucleus is of great interest for such an experimental study, as its first isomeric I$_f^{\pi}$ = 9/2$^-$ state lies at 6.2 keV above the I$_i^{\pi}$ = 7/2$^+$ stable ground state which is of the order of magnitude of the electron energies of interest to probe $(e,e')$ cross sections. This nuclear excited state has a half life of 6.05 $\mu$s which is long enough to allow its detection. Study of the $^{181}$Ta(e,e')$^{181}$Ta$^*$ process at energies of a few tens of keV requires very high intensity electron pulses with duration lower than 100 ns in order to decouple the excitation step from the nuclear de-excitation one. The number of electrons in a pulse and the repetition rate must be high enough to achieve a reasonable cumulative excitation yield in a limited experimental time. We have recently developed a 10 Hz electron plasma source capable of delivering 10$^{14}$ electrons per bunch with kinetic energies up to 30 keV\cite{gobet2020versatile}. In this paper, we evaluate the possibilities opened with this new device to constrain the models describing the $(e,e')$ process at low energy.

This paper is organized as follows. In section II, the different models developed to calculate the $(e,e')$ cross sections at low energy will be briefly described. The cross sections will be presented for $^{181}$Ta considering Coulomb potentials of a bare nucleus (unscreened potential) or of a nucleus in an atom (screened potential). In section III, the characteristics of our intense, pulsed multi-keV electron source will be described. They will be used to quantify, in a simple scheme of tantalum target irradiation, the expected values of the nuclear excitation yields with the different $(e,e')$ models. In section IV, we will identify the most relevant experimental observable to measure the excitation yield. Section V will be dedicated to specific effects of this innovative electron source that could modify the nuclear excitation yield or the probabilities of detecting the experimental observable. Finally we will conclude on the experimental difficulties that could be encountered to constrain the models describing the $(e,e')$ process in the keV - tens of keV energy range.

\section{ Theoretical description of the \MakeLowercase{(e,e')} process in unscreened or screened Coulomb nuclear field for $^{181}$Ta}

\subsection{Semi-classical approach for nuclear Coulomb excitation in a bare nucleus}

The first $(e,e')$ cross section calculations in $^{181}$Ta at low energy were performed in 1994 by Lethokov and Yukov considering a classical treatment of the Coulomb excitation process\cite{letokhov1994excitation}. In this approach, the electron is moving along a hyperbolic orbit in the attractive
Coulomb field of the tantalum bare nucleus with a differential scattering cross section
given by the classical Rutherford law. Assuming that the trajectory of the electron is not appreciably affected by the nuclear excitation, the 
differential nuclear excitation cross section in a given direction is obtained by multiplying the Rutherford cross section by the probability $P$ for the nucleus to be excited from the initial state $i$ to the final state $f$ along the electron trajectory\cite{alder1956study}. This probability is
\begin{equation}
P=(2I_i+1)^{-1} \sum_{M_i,M_f} |b_{if}|^2
\end{equation}
where  $M_i$, $M_f$ are the magnetic quantum numbers of the initial and final nuclear states and $b_{if}$ the amplitude of the transition. 
This transition amplitude can be expressed for electric excitation and under the first-order time dependent perturbation approximation as
\begin{equation}
b_{if}=\frac{1}{i\hbar}\int_{-\infty}^{\infty}<f\mid\mathcal{H}_E(t)\mid i>e^{i\omega t} dt
\end{equation}
where $\mathcal{H}_E(t)$ is the Coulomb energy along the electron trajectory leading to the given final direction and
\begin{equation}
\omega = \frac{\Delta E}{\hbar}=\frac{E_f-E_i}{\hbar}
\end{equation}
is the nuclear frequency associated to the excitation energy $\Delta$E = 6.2 keV.

As demonstrated in ref.\cite{alder1956study}, the transition amplitude of eq.(2) can be expressed in terms of the reduced nuclear transition probability B(E1) = 2$\times 10^{-6}$ of the electric-dipole transition in a $^{181}$Ta nucleus\cite{wunucl}.
The total excitation cross section is obtained after an integration over all scattering directions and therefore all impact parameters. The calculated cross sections presented in Fig.1 (Semi classical - Bare nucleus) show a linear evolution in logarithmic scale signing a power law which decreases with increasing energy over the energy range 7-100 keV. These cross sections are of the order of a few 10$^{-31}$ cm$^2$ at the excitation threshold and a few 10$^{-32}$ cm$^2$ above 100 keV.

The classical description of the inelastic collision is highly questionable as the energy loss of the electron during the nuclear transition is not small compared to the incident energy, so the effect of the nuclear excitation on the particle motion cannot be neglected. 

%Fig.~\ref{fig:cross_sec}
\begin{figure}
\includegraphics[height=8cm, trim=0 40 220 160, clip=true]{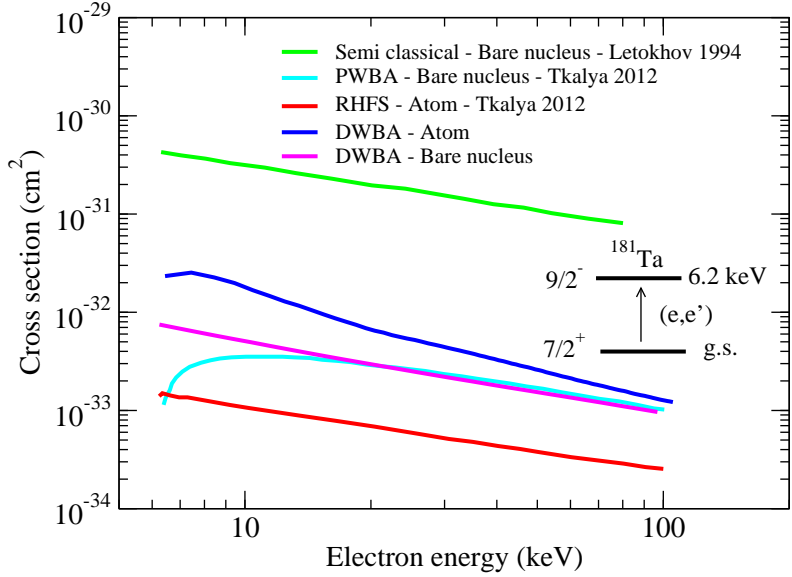}% Here is how to import EPS art
\caption{\label{fig:cross_sec} $^{181}$Ta electron inelastic excitation cross sections in semi-classical, PWBA, DWBA and RHFS methods in screened or unscreened Coulomb nuclear field (Letokhov 1994: \cite{letokhov1994excitation}, Tkalya 2012:  \cite{tkalya2012cross}).}
\end{figure}

\subsection{Quantum-mechanical approaches in unscreened or screened  Coulomb nuclear field}

More complete quantum-mechanical treatments of inelastic electron scattering have been performed by several authors\cite{gosselin2009nuclear,dzyublik2013role,tkalya2012cross} in the multi-keV energy range.
In these approaches, the interaction was treated as a static Coulomb field \cite{gosselin2009nuclear,dzyublik2013role} or as a result of exchange by virtual photons between the electrons and the nucleus \cite{tkalya2012cross}. The excitation cross sections can be expressed as sums of terms referring to the different orbital momentum $l_i$ and $l_f$  of the incoming and outgoing electron respectively. Each of the terms involves the reduced transition probability B(E1) and a radial matrix element $R_{l_i l_f}$ of the form

\begin{equation}
R_{l_il_f}= \frac{1}{k_ik_f} \int_{0}^{\infty} \frac{F_{l_i}(k_ir)F_{l_f}(k_fr)}{r^2}dr
\end{equation}

where $r$ is the distance from the point-like nucleus, $k_i$ and $k_f$ are the wave numbers of the incoming and outgoing electron and $F_l(kr)$ is the radial part of the electron wave functions decomposed into partial waves, which are solutions of the Schr\"{o}dinger or Dirac equation for an unscreened or screened Coulomb potential. The description of the wave functions describing the projectile is the key point of the cross section calculation.

Using plane wave functions to describe the electron before and after the collision is the easiest way to quantify the radial matrix elements. The  Plane Wave Born Approximation (PWBA) model is justified if the collision time is short compared to the time needed to exchange a virtual photon between the electron and the nucleus. The PWBA calculations of Tkalya et al \cite{tkalya2012cross} in a bare $^{181}$Ta nucleus are presented in Fig.1 (PWBA - Bare nucleus) and predict a cross-section of a few 10$^{-33}$ cm$^2$ for 7-100 keV electrons. This is 100 times lower than the values obtained with the semi-classical method.

The Distorted Wave Born Approximation (DWBA) model allows to go further in the description of the electronic wave function. Gosselin et al have estimated the wave function distortion by solving the Schr\"{o}dinger equation with an unscreened Coulombian interaction potential. They predicted $(e,e')$ cross section values of the order of 10$^{-31}$ - 10$^{-32}$ cm$^2$ for the transition to the first nuclear state in $^{110}$Ag and $^{201}$Hg nuclei over an energy range of up to 100 keV\cite{gosselin2009nuclear}. The $(e,e')$ cross sections have been calculated on $^{181}$Ta by the authors for the present study. They are of the order of 10$^{-33}$ - 10$^{-32}$ cm$^2$ as reported in Fig.1 (DWBA - Bare nucleus). The results of DWBA and PWBA calculations converge for electron energies above about 20 keV: below this energy, we cannot neglect the deformations of the wave functions describing the incoming and outgoing electron.

Taking into account the effect of screening of the nuclear field by bound atomic electrons on the $(e,e')$ cross sections is a very complex task that has been investigated by a few authors\cite{dzyublik2013role,tkalya2012cross}.
For example, Dzyublik et al.\cite{dzyublik2013role} considered a DWBA model in which the screening of the Coulomb field of the nucleus by the bound atomic electrons is modeled by a potential of the form

\begin{equation}
V_c(r) = -\frac{Ze^2}{r} e^{-r/r_o}
\end{equation}

where $r_o$ is a characteristic screening length and $Ze$ is the electric charge of the $^{181}$Ta nucleus: the longer the length $r_o$, the weaker the screening. The limit of the bare nucleus is obtained for $r_o \rightarrow +\infty$.
In the case of $^{181}$Ta, the screening length is considered equal to 0.013 nm, which corresponds to the characteristic radius of a tantalum atom predicted by the Thomas-Fermi model. In this study, the wave functions of the incident electron have been calculated considering this screened potential. 
The $(e,e')$ cross sections obtained by this method are reported in Fig.1 (DWBA - Atom). They are of the order of a few 10$^{-32}$ cm$^2$. For electron energies about 10 keV, they are 3 to 4 times higher than those obtained for a bare nucleus with the DWBA method. Moreover, these calculations suggest that effects of screening are not significant for electron energies higher than 60 keV.

The relativistic Hartree-Fock-Slater (RHFS) method is another way to describe the screening effects within the atom. It allows to compute in a self-consistent way the bound atomic electronic wave functions and the mean interaction potential between the nucleus and the electrons by solving the Dirac equation. This same equation is also used to study, in a second step, the interaction between an incident electron and the tantalum atom. Tkalya et al. established RHFS calculations\cite{tkalya2012cross} for both $^{181}$Ta nucleus in a neutral atom and in an ion with a degree of ionization of 33+. The authors showed that the $(e,e')$ nuclear excitation cross sections are the same for the two systems and that the electrons of the outer layers do not play a major role in the nuclear Coulomb potential screening.
In this model, $(e,e')$ cross sections are expected to be of the order of 10$^{-34}$ - 10$^{-33}$ cm$^2$ over the energy range 7 - 100 keV (see Fig.1 / RHFS - Atom). These values are one order of magnitude lower than those obtained with the screened potential DWBA method and do not converge towards the PWBA or unscreened DWBA results for high energy electrons. 

It is difficult, through physical sense, to gauge the effect of potential screening on the $(e,e')$ cross sections. In a bare nucleus, the kinetic energy of the electrons in the vicinity of the nucleus is greater than its initial value because of the nuclear Coulomb attraction. In an atom, the electronic cloud will decrease the range of the effective attractive potential well. As the incident electrons will be much less attracted by the nucleus, their kinetic energy  in the vicinity of the nucleus will be closer to their initial value. In that case, incident electrons spend more time in the vicinity of the nucleus, which could increase the probability of Coulomb excitation. On the other hand, the intensity of the interaction decreases between the electron and the screened nucleus, which may result in a lower nuclear excitation probability. Determining whether one of the two effects prevails is a complex issue. In the absence of theoretical arguments allowing to definitely conclude on the screening effects, experimental results are needed to constrain these models.  In the following, we propose an original experimental scheme for such a challenging measurement.

\section{\MakeLowercase{(e,e')} and \MakeLowercase{($\gamma$,$\gamma$')} excitation yields in a $^{181}$Ta target irradiated with a High-intensity electron beam from a biased laser-plasma}

The main characteristics of our newly developed electron source were already carefully described\cite{gobet2020versatile,raymond2017energy,versteegen2019role,comet2016absolute} and are briefly summarized in the following. A schematic view of the device is displayed in Fig.2(a). A 10 ns, 10$^{13}$ W cm$^{-2}$ Nd:YAG laser pulse is focused on an aluminum target at a repetition rate up to 10 Hz. Each shot produces a plasma in which about 2$\times$10$^{15}$ electrons are free. This plasma presents two components: a dense aluminum plasma, with densities reaching 10$^{20}$-10$^{22}$ part.m$^{-3}$, preceded by a Low Density anisotropic Pre-Plasma (LDPP), containing approximately 10$^{16}$-10$^{17}$ part.m$^{-3}$. Those two-components expand during 130 ns between the aluminum  target, biased at a negative voltage -$V_T$, and a thick anode plate located 50 mm downstream from the target. 

Electrons are extracted and accelerated from the front end boundary of the dense plasma component biased at the effective potential $-V_p$ which acts as a moving plasma cathode\cite{raymond2017energy}. In Fig.2(b), we have reported the measured energy distributions of the electrons that have reached the 24.6 cm$^2$ central area of the anode in a duration of about 50-70 ns for 5 target voltages. The number of incident electrons slightly increases from 7$\times$10$^{13}$ at $V_T$=10 kV to 9$\times$10$^{13}$ at $V_T$=30 kV. These distributions are continuous and indicate a maximum energy greater than $eV_T$. Indeed, the first extracted electrons are accelerated toward the anode by the electric field induced by $V_T$ but they are also pushed by their followers, allowing them to gain additional kinetic energy. Depending on the target voltage, the average energy of the extracted electron beam varies from $\approx$ 2 keV to $\approx$ 8 keV. At these energies, the cross sections of the two models accounting for Coulomb potential screening differ by a factor of 15 to 20 (see Fig.1).

\begin{figure}
\begin{minipage}[h]{0.5\linewidth}
\includegraphics[height=5.cm,trim=0 0 0 0, clip=true]{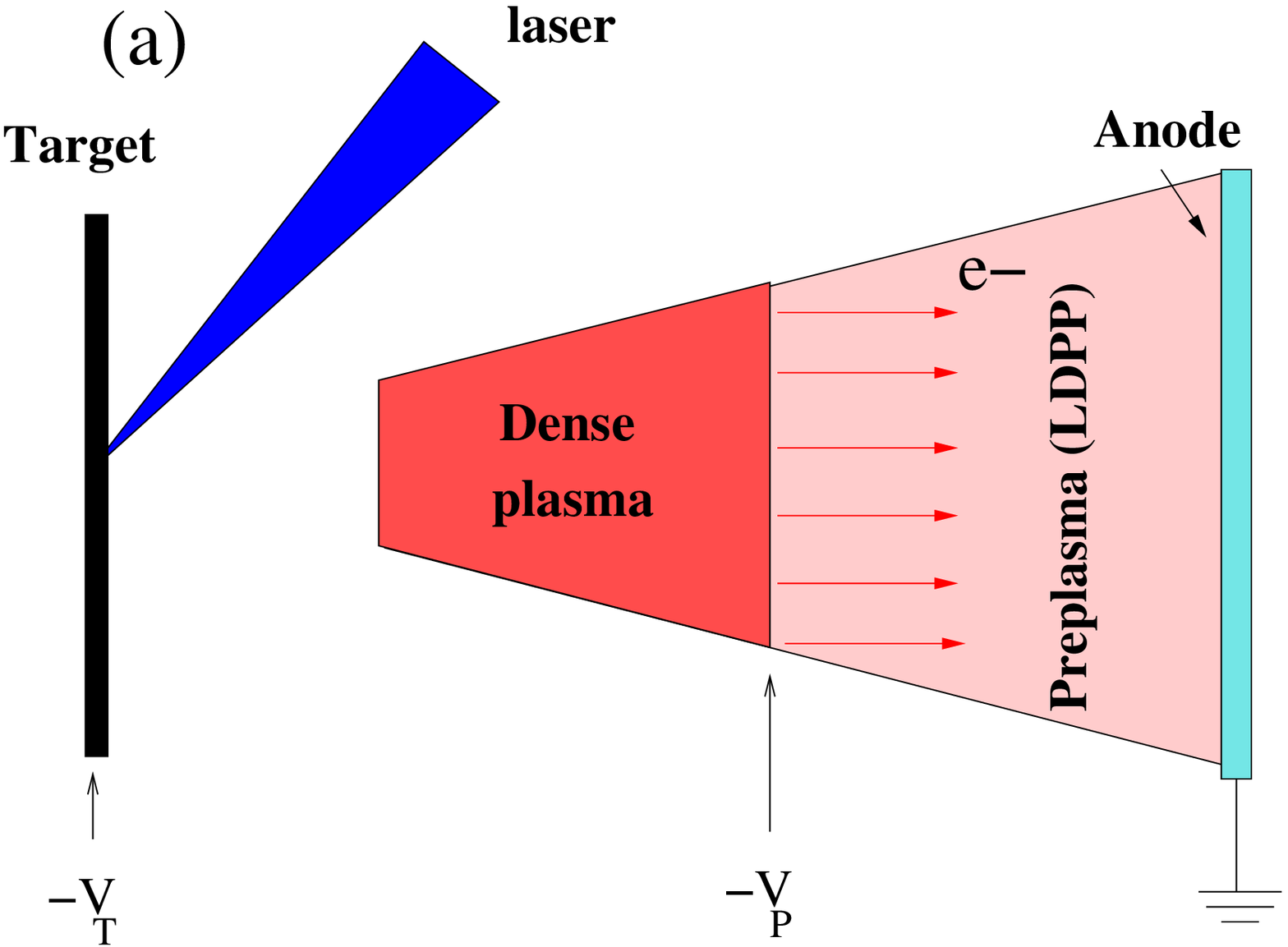}
\end{minipage}
\begin{minipage}[h]{0.4\linewidth}
\includegraphics[height=7.cm,trim=0 40 200 150, clip=true]{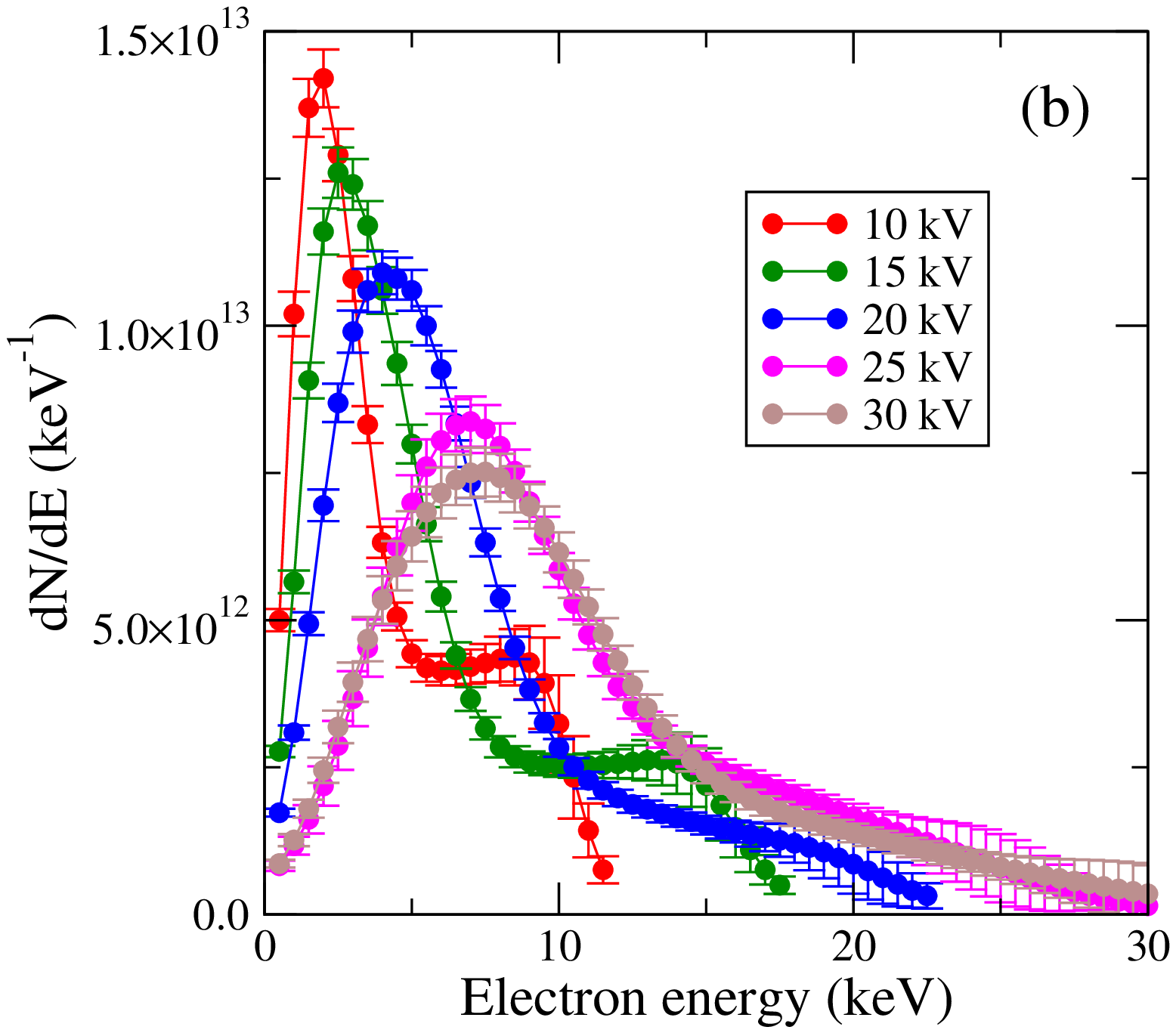}
\end{minipage}
\caption{(a) Schematic of the compact high-intensity electron source. The plasma positions are indicated some time after the laser shot. (b) Energy distributions of the electrons impinging the 24.6 cm$^2$ anode at various target voltage V$_T$.}
\end{figure}

Using a tantalum anode would be the simplest way to use this source as an irradiation
facility to investigate $(e,e')$ nuclear excitation of $^{181}$Ta. In the following we will calculate the nuclear excitation yields that could be obtained by irradiating this tantalum anode with these electron beams. Although the cross sections of interest for this type of experiment are those considering a screened potential, we have generalized the calculations for all the models. This allows us to get orders of magnitude of expected yields for different values of cross sections. Moreover, we will present some results in the specific case of the DWBA model on a bare nucleus insofar the cross sections of this model are intermediate to those of the two screened potential models at energies lower than 20 keV.
The nuclear excitation yields in such a
sample can be calculated using the Monte Carlo code Geant4 with its 10.4.p01 version\cite{agostinelli2003geant4}
associated with the Livermore Physics List\cite{allison2016recent}. This allowed us to compute
particle interactions (electrons, photons) with energies as low as
250 eV. Multiple scattering, Bremsstrahlung, and continuous energy
loss are taken into account for electrons. The physical processes for
photons are the Compton and photoelectric effects and Rayleigh scattering.
In addition, secondary electron emission and fluorescence are
also included. In the following numerical studies, all material characteristics come from the National Institute Standards and Technology material lists. The electron energy distributions reported in Fig.2(b) are used as input and 10$^8$ simulated electrons interacting with the Ta anode are a reasonable trade-off to save calculation time while maintaining the statistical error below 5\%. The electron beam impinges on the tantalum anode perpendicular to the surface. This anode has a thickness of 3 $\mu$m which is larger than the range of 30 keV electrons in tantalum\cite{estar}.

The inelastic electron scattering $(e,e')$ and the photoexcitation ($\gamma$,$\gamma$') nuclear excitation processes are integrated in the code. For the $(e,e')$ mechanism, the simulations are performed by fixing one of the previously discussed models and multiplying the calculated cross sections by a factor 10$^{10}$ to ensure a good statistical accuracy. With this amplification factor, the effective cross sections are of the order of 10$^{-24}$ - 10$^{-21}$ cm$^2$ in the code. The process remains negligible compared to other particle-matter mechanisms. The numbers of excited nuclei reported in this paper are corrected from this factor.

Unlike the electron inelastic scattering cross sections, the photoexcitation cross section is very high ($\sigma_{ph.ex}$= 1.7$\times$10$^{-18}$ cm$^2$)\cite{frauenfelder1962mossbauer}. However, the process is strongly resonant as the width of the $^{181}$Ta isomeric state is very small ($\Gamma$ = 6.7$\times$10$^{-11}$ eV). To evaluate the photoexcitation yield, we have counted the number of 6.237 $\pm$ 0.050 keV X-ray photons created during the slowing down of the electrons inside the tantalum target followed by their absorption via a photoelectric process. At this photon energy, the cross section of this process is $\sigma_{photoel.}$= 9$\times$ 10$^{-20}$ cm$^2$. The final number of photoexcited nuclei in the interaction is obtained by dividing this output by the energy width (100 eV) of the X-ray photon energy range and multiplying it by the energy width $\Gamma$ of the tantalum isomeric state as well as by the cross section ratio $\sigma_{ph.ex}$/$\sigma_{photoel.}$. 
Finally, a normalization factor is applied to the calculated nuclear yields to take the measured number of incident electrons into account. 

The nuclear excitation yields per laser shot are plotted in Fig.3 as a function of the target voltages for all the excitation models previously described.
In all cases the number of excited nuclei per laser shot increases by a factor 10 for target voltages rising from 10 to 30 kV. The excitation yields vary from a few 10$^{-2}$ to a few 10 excitations per shot depending on the different excitation models $(e,e')$ and the voltages. The dependence of the excitation rate (e,e') on the target voltage slightly differs  for  PWBA model on a bare nucleus. As reported in Figure 1, the effective cross section associated with this model shows a strong deviation from a power law for electrons of energy lower than 20 keV compared to the other models. 
Moreover, in this type of experiment, the $^{181}$Ta nuclei would be mostly excited by the $(e,e')$ process whatever the model. Indeed, the conversion of incident electron energy into X-ray photons by Bremstrahlung is not efficient in this energy range. The low X-ray photon number coupled with a very resonant excitation mechanism makes the photoexcitation process very unlikely.

\begin{figure}
\includegraphics[height=8cm, trim=0 50 30 140, clip=true]{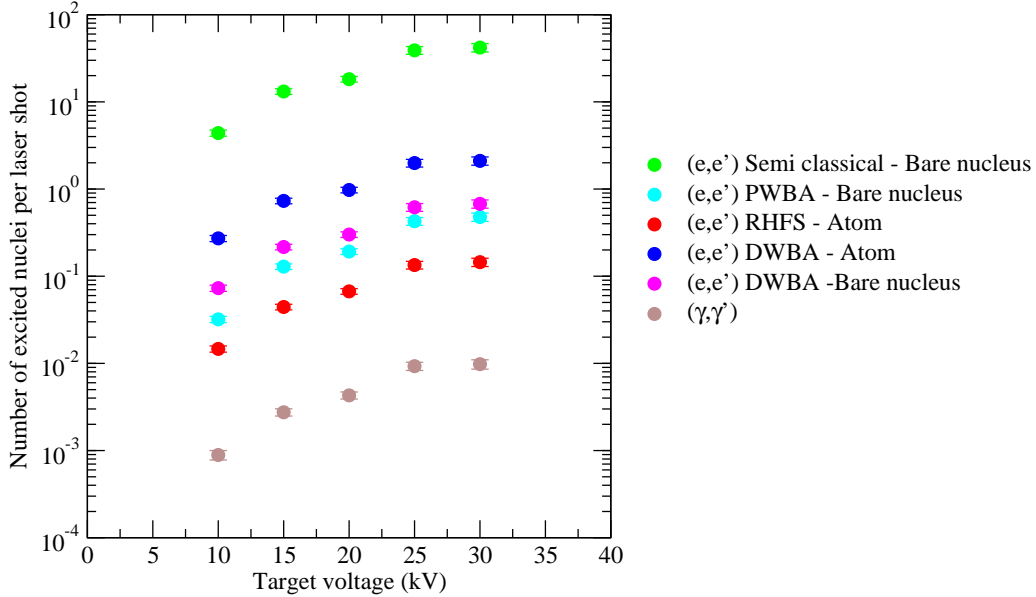}% Here is how to import EPS art
\caption{\label{fig:cross_sec} $^{181}$Ta excitation yields per laser shot via photoexcitation and (e,e') processes. For (e,e') the cross sections of Fig.1 have been considered.}
\end{figure}

As we will see in the next section, the depth distribution of excited nuclei is one of the key parameters to estimate the production yields of experimental observables. Incident electrons enter through the front face of the tantalum anode (depth $x=0$) as shown in the inset of Fig.4(a). The number of excited nuclei in the tantalum anode is recorded as a function of its depth with a resolution of 1 nm and is cumulated over 10 000 laser shots to facilitate reading of the graph. The depth distributions calculated with the unscreened potential DWBA model are shown in Fig.4(a) for the different target voltages. As previously written, $(e,e')$ cross sections of this model have intermediate values to those obtained with the screened potential models. 
The excitation yields calculated with the DWBA model with screened potential are about 3 to 4 times higher while those obtained with the RHFS model are 5 to 6 times lower. The depth distribution reported in Fig.4(a) is decreasing over a characteristic distance up to 500 nm at 30 kV, as expected given the range of electrons in tantalum at these kinetic energies. The integrated $(e,e')$ excitation yield over the tantalum depth reported in Fig.4(b) shows that most of the nuclear excitations would occur in the first 300 - 400 nanometers of the tantalum anode face. Calculations have been carried out for all the $(e,e')$ models and show similar dependencies.

\begin{figure}
\includegraphics[height=8cm, trim=0 50 120 250, clip=true]{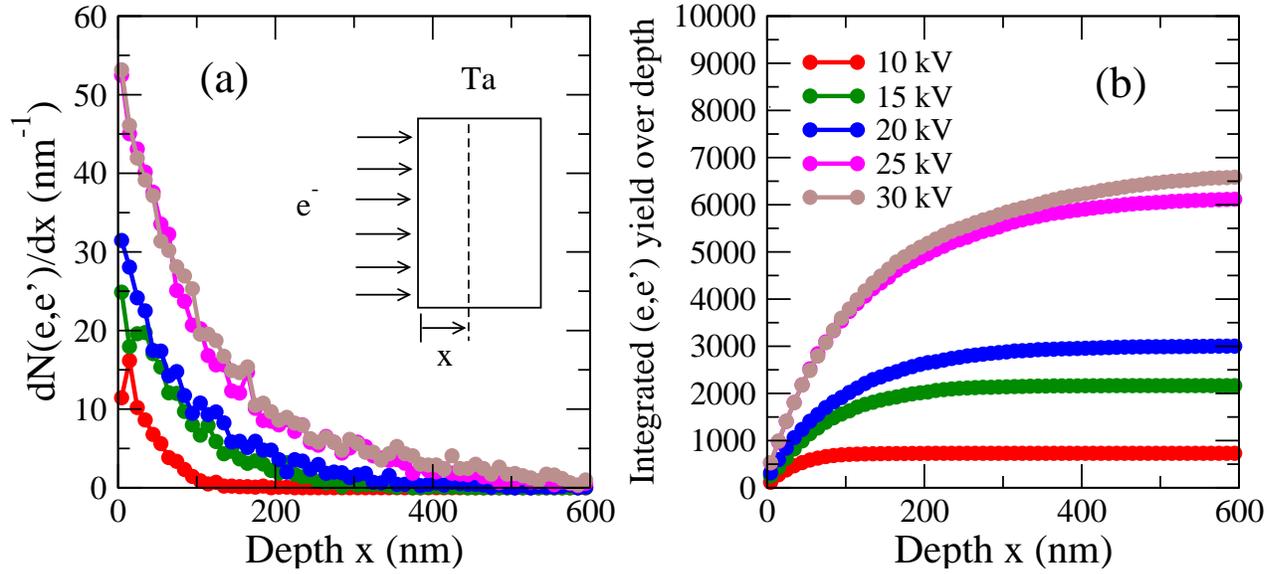}% Here is how to import EPS art
\caption{\label{fig:cross_sec} (a) $^{181}$Ta excitation yield profiles along the tantalum anode depth. DWBA (e,e') cross sections of a bare nucleus have been considered. Excitation yield profiles are calculated considering a statistics of 10 000 laser shots. (b) Integrated $^{181}$Ta excitation yield profiles along the tantalum anode depth.}
\end{figure}

\section{The $^{181}$T\MakeLowercase{a}$^*$ decay observables}

The first excited nuclear state of $^{181}$Ta de-excites either by internal conversion or by $\gamma$ emission. The internal conversion coefficient is measured in a neutral atom with a value of $\alpha$(E1) = 70 which makes conversion electron emission the most probable path\cite{wunucl}. In this section we will study the emission yields of $\gamma$ photons and conversion electrons in order to identify the most relevant experimental observable to study the (e,e') process in neutral atoms.

\subsection{$\gamma$-ray photons}

In a first step, we have calculated the yields of $\gamma$-ray photons exiting the anode for the different electron inelastic scattering models and the target voltages considered in this study. In order to optimize the solid angle of detection, it is more relevant to consider a scintillator or a semiconductor detector with a large collecting surface placed at the back of the anode. The absorption length of the 6 keV photons being about 2 $\mu$m in tantalum it is important that the tantalum anode is not too thick ($<$ 1 $\mu$m) in order to avoid self-attenuation.
Such a very thin anode could be weakened during a laser shot. Then, as reported in the inset of Fig.5, we will consider a tantalum layer of thickness $x$ deposited on a thicker structure of light Z atomic number such as a 15 $\mu$m thick aluminum sheet for which we have experimentally verified the good resistance on several hundreds of laser shots.

\begin{figure}
\includegraphics[height=8cm, trim=0 50 310 170, clip=true]{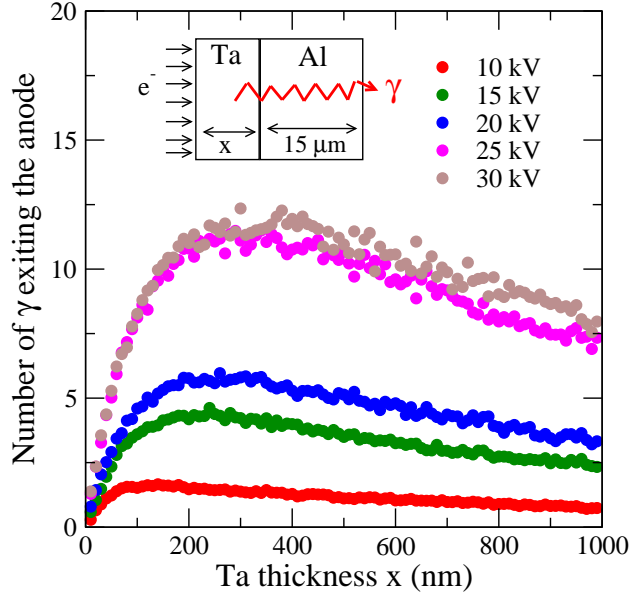}% Here is how to import EPS art
\caption{\label{fig:cross_sec} Number of $\gamma$-ray photons exiting the anode as a function of the tantalum anode thickness for various target voltages. (e,e') nuclear excitation profiles have been considered using DWBA (e,e') cross sections of a bare nucleus. The number of exiting photons are calculated considering a statistics of 10 000 laser shots.}
\end{figure}

We have calculated, by a Monte Carlo method, the number of $\gamma$-ray photons exiting the anode back side considering the depth distribution of emission of these photons following the curves of Fig.4(a). For each depth, 10$^4$ photons are emitted and the number of exiting $\gamma$-ray photons is counted. This output is normalized to the number of excited nuclei corrected from the total conversion coefficient and cumulated over 10 000 laser shots. The calculation is carried out for different thicknesses $x$ of the tantalum layer in order to determine its optimal value. The expected number of exiting photons is plotted in Fig. 5 as a function of the thickness $x$ for the different target voltages and considering the excitation (e,e') cross sections described by the DWBA model with an unscreened potential.
Whatever the target voltage, the curves show a maximum number of detectable $\gamma$-ray photons  for a tantalum thickness of about 200 to 400 nm. This optimum is also reached for all the $(e,e')$ models presented in section II. For smaller thicknesses, the number of detectable photons is limited by the number of excited nuclei in the anode.  For larger thicknesses, the number of excited nuclei remains constant but the self-attenuation of photons in the tantalum layer or the aluminum support becomes limiting. The number of detectable $\gamma$-ray photons cumulated over 10 000 laser shots is reported in Fig.6 for the thickness of the tantalum layer set at the optimal value of 300 nm. These photon yields increase with the target voltage for all the $(e,e')$ cross section models. Nevertheless they remain low and vary between 3 and 800 detectable photons depending on the models at 30 kV. A difference of a factor about 15 is expected between the $\gamma$-ray photon yields calculated with the DWBA ($\approx$ 40 $\gamma$-ray photons) and RHFS ($\approx$ 3 $\gamma$-ray photons) models with screened Coulomb potential.

\begin{figure}
\includegraphics[height=8cm, trim=0 50 320 130, clip=true]{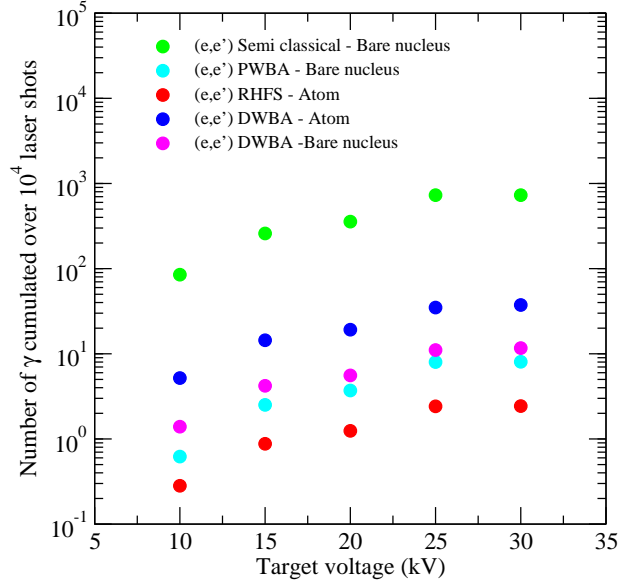}% Here is how to import EPS art
\caption{\label{fig:cross_sec} Number of $\gamma$-ray photons exiting the anode as a function of the target voltage considering a 300 nm thick Ta anode. $(e,e')$ nuclear excitation profiles have been considered using cross sections of models described in Section II. The number of exiting photons are calculated considering a statistics of 10 000 laser shots.}
\end{figure}

Given these very low numbers it seems very complex to rely on this experimental observable to constrain the nuclear $(e,e')$ excitation models with a screened Coulomb potential. The experiment is even more complicated as a strong flash of X-ray photons will be emitted during the slowing down of the incident electrons in the anode over a duration of a few tens of ns. In a previous study\cite{gobet2021x}, we have shown that the total energy of the photons of this flash was of the order of 10$^9$ to 10$^{10}$ keV in the backside of a 15 $\mu$m thick aluminum anode. Such an amount of energy cannot totally relax in a scintillator or semiconductor detector\cite{tarisien2018scintillators} over a few tens of $\mu$s, i.e. a duration of the order of the half-life of the first excited state in $^{181}$Ta (6.05 $\mu$s). 

\subsection{Internal conversion electrons}

In an atom, the first excited state of the $^{181}$Ta nucleus mainly decays by internal conversion with electrons of kinetic energies below 6.2 keV. The electron path in tantalum in this energy range is less than 100 nm. Given the depth profile in Fig.4(a), these electrons have a higher exit probability through the front side of the anode. By applying an adequate pulsed electric field a few 10 $\mu$s after the laser shot, electrons exiting from the front side of the anode can be extracted and post-accelerated towards a detector shielded from the radiation generated by the laser-target or electron beam-anode interactions. In this section we will estimate the number of conversion electrons that can exit the anode considering the different (e,e') excitation models  and the different target voltages.

The conversion coefficients $\alpha$(E1) for the E1 transition in the first excited level of $^{181}$Ta were calculated using Dirac-Hartree-Fock-Slater wavefunctions in a neutral atom with the code CATAR\cite{pauli1975computer}. The values of $\alpha$(E1), the probabilities of electron emission a well as the experimental and theoretical kinetic energies of the emitted electrons are shown in Table I.
Note that the total theoretical internal conversion coefficient is about 30 while a value equal to 70 is measured experimentally. Dynamic effects in the nucleus volume \cite{church1960nuclear,green1958nuclear,kramer1962penetration} are not taken into account in the CATAR code. They can probably explain the difference between the calculated and measured internal coefficients. These effects may lead to an increase in the conversion probability for electrons that are most bound to the nucleus. As electron emission probability has not been experimentally measured we have considered the theoretical values to estimate the number of conversion electrons which can leave the target after a laser shot.

\begin{table}[h]%The best place to locate the table environment is directly after its first reference in text
\caption{\label{tab:table1}%
Theoretical internal conversion coefficients and probability of electron emission for E1 transition calculated using the code CATAR\cite{pauli1975computer} for the 6.2 keV transition in $^{181}$Ta. Theoretical values of conversion electron kinetic energies are also given and compared with the experimental values.
}
\begin{ruledtabular}
\begin{tabular}{ccccc}
\textrm{Orbital shell}&
\textrm{$\alpha$ (E1)}&
\textrm{Probability (\%)}&
\textrm{E$_{el. th.}$ (keV)}&
\textrm{E$_{el. exp.}$ (keV)}\\
\colrule
M1 & 2.11 & 7 & 3.615 & 3.533\\
M2 & 3.40 & 11.3 & 3.832 & 3.772\\
M3 & 6.61 & 22.0 & 4.111 & 4.127\\
M4 & 5.39 & 18.0 & 4.517 & 4.448\\
M5 & 7.47 & 24.9 & 4.579 & 4.506\\
N1 & 0.57 & 1.9 & 5.688 & 5.675\\
N2 & 0.76 & 2.5 & 5.779 & 5.776\\
N3 & 1.42 & 4.7 & 5.841 & 5.836\\
N4 & 0.91 & 3.0 & 5.999 & 5.999\\
N5 & 1.26 & 4.2 & 6.010 & 6.011\\
N6 & 0.01 & 0.1 & 6.206 & 6.213\\
N7 & 0.12 & 0.4 & 6.207 & 6.215\\
\end{tabular}
\end{ruledtabular}
\end{table}

We have calculated, using the code Geant4, the probability of a conversion electrons to exit the anode front side as a function of its emission depth. The calculations are carried out for electron emission depths ranging from 5 to 95 nm in 5 nm steps. For each configuration, 10$^4$ electrons are randomly emitted at a given depth taking into account the emission probabilities and the theoretical kinetic energies of the conversion electrons given in Table I. The electrons that exit through the front side of the anode are counted. Dividing this output by the number of emitted electrons in the code and correcting by the value of the total conversion coefficient, we obtain the probability that the de-excitation of a $^{181}$Ta nucleus will lead to the emission of an electron from the anode front side. This probability, reported in Fig.7, falls with the depth of the excited Ta nucleus: it is lower than 0.1 for depths greater than 50 nm. In this context, only Ta nuclei excited in the first 30 nm of the anode can emit conversion electrons detectable from its front surface.

\begin{figure}
\includegraphics[height=8cm, trim=0 50 300 130, clip=true]{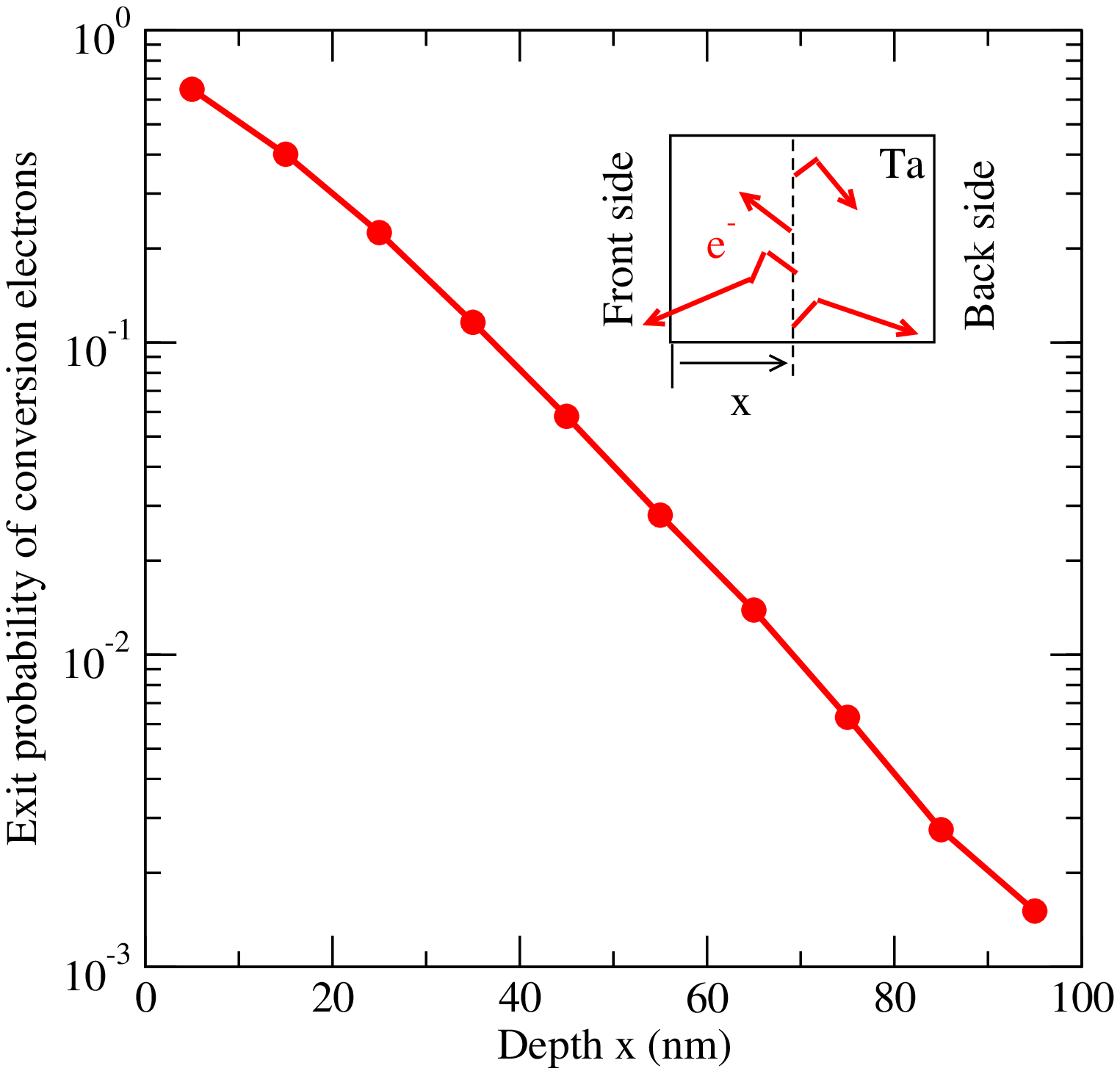}% Here is how to import EPS art
\caption{\label{fig:cross_sec} Probability that a conversion electron exits through the front side of the tantalum anode as a function of its emission depth x.}
\end{figure}

The number of conversion electrons that can exit this surface is obtained by integrating the depth distributions of the excited tantalum nuclei and weighted by the exit probabilities presented in Fig.7. The results of these calculations performed for all $(e,e')$ excitation models  and all target voltages are reported in Fig.8 for a statistics of 10 000 laser shots. On average, about 100 times more conversion electrons are emitted from the front side compared to the number of $\gamma$-ray photons detectable on the back side of the anode in the optimal geometrical configuration. At 30 kV, there are 20 times more electrons expected in the DWBA screened model ($\approx$ 2 000 electrons) compared to the values obtained with the RHFS one ($\approx$ 100 electrons).

\begin{figure}
\includegraphics[height=8cm, trim=0 50 300 130, clip=true]{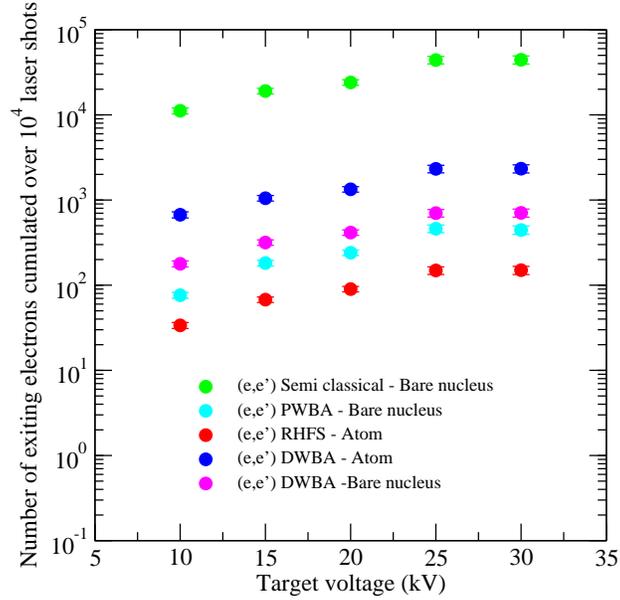}% Here is how to import EPS art
\caption{\label{fig:cross_sec} Number of conversion electrons exiting the front side of the anode as a function of the target voltage. (e,e') nuclear excitation profiles have been considered using cross sections of models described in Section II. The number of exiting electrons are calculated considering a statistic of 10 000 laser shots.}
\end{figure}

Designing an experiment to measure excitation yields requires knowledge of the energy distribution of electrons exiting the anode. Although electrons are emitted from nuclei at well-defined energies as shown in Table 1, these energies at the anode exit can be degraded due to electron interaction processes with tantalum. These distributions were calculated for target voltages of 10 and 30 kV with the Geant4 code considering the depth distributions of excited nuclei reported in Fig. 4(a). The calculated energy distributions shown in Fig. 9 are continuous and slightly degraded due to the partial slowing down of electrons in the anode. The experimental electron extraction setup will therefore need to rely on electrodes biased at least to 10 - 20 kV to guide the particles to a shielded detector. The target voltage will also have to be inhibited during the extraction time in order not to disturb the electron extraction.

\begin{figure}
\includegraphics[height=8cm, trim=20 50 400 220, clip=true]{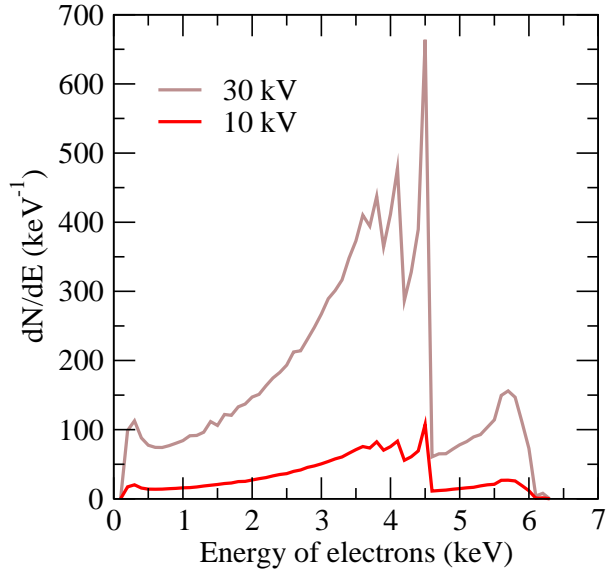}% Here is how to import EPS art
\caption{\label{fig:cross_sec} Energy distributions of the conversion electrons exiting the front side of the anode considering the nuclear excitation profiles of Fig.4(a) at target voltages of 10 and 30 kV. The energy distributions are calculated considering a statistics of 10 000 laser shots.}
\end{figure}

\section{Plasma effects on the tantalum target}

The previous analyses show that the measurement of the number of excited $^{181}$Ta nuclei per $(e,e')$ process is possible through the detection of conversion electrons. In the following, we will examine the effects of the plasma and the intense electron beam on the tantalum anode. The energy deposition by the electron beam on the anode could induce hydrodynamic effects and material loss. Moreover the aluminum plasma is deposited shot after shot on the anode. This layer could have two effects: a modification in the nuclear excitation yield and a decrease of the probability of the conversion electrons to leave the anode. 

\subsection{Effects of electron heating}

The interaction of the electron beam delivered by the device with a tantalum anode is a succession of complex processes. The deposited beam energy is converted into internal energy and can lead to a local rise in temperature and pressure. We have computed the depth dose profile in tantalum for different target voltages. For each configuration, 1 billion electrons are sent perpendicular to the sample with the energy distributions reported in Fig.2(b). The energy deposited by the primary and the secondary particles in the sample is recorded as a function of depth, with a resolution of 1 nm. The dose profile is obtained by dividing the total energy deposited in each 1 nm-thick disk by its mass. Then, it is rescaled to the experimental number of electrons measured in the 24.6 cm$^2$ area as reported in Fig.2(b). The results of these simulations are shown in Fig.10. The curves represent the dose profile D(x) for the six values of target voltage. A maximum dose of D$_{max} \approx$ 4 - 6 kJ kg$^{-1}$ is reached in each laser shot whatever the target voltage in a characteristic length $\delta \approx$ 100-300 nm. Using the specific heat C$_v \approx$ 140 J K$^{-1}$ kg$^{-1}$ of tantalum material, we can then estimate the anode temperature increase $\Delta$T$\sim \frac{\Delta E}{C_v} \approx$ 30 - 40 K. The resulting temperature is much lower than the melting temperature of tantalum which is about 3200 K. If the energy deposition is isochoric, the resulting pressure increase would be estimated by using the relation $\Delta P \sim \Gamma.\rho.D_{max} \approx$ 0.1 GPa $\approx$ 1 bar where $\Gamma \approx$ 1.6 is the Grüneisen coefficient of tantalum\cite{meyers1994dynamic}. Although these numbers give only orders of magnitude, they show that a significant increase of pressure and temperature could be expected in the experiment.

\begin{figure}
\includegraphics[height=8cm, trim=50 50 300 170, clip=true]{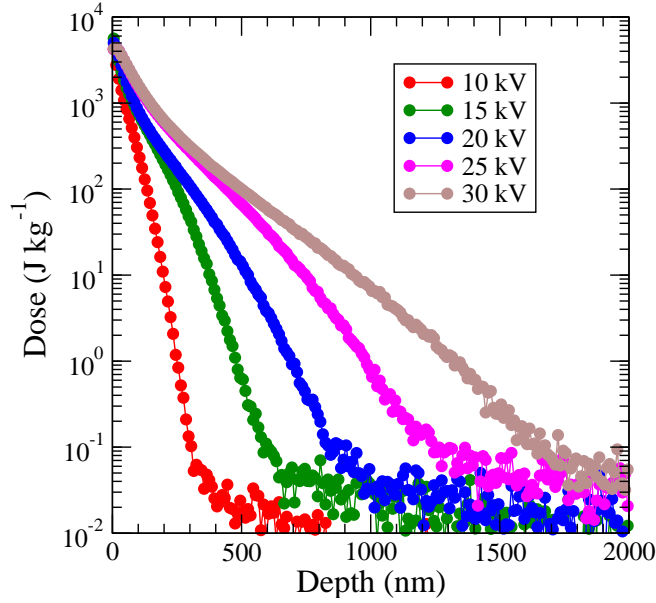}% Here is how to import EPS art
\caption{\label{fig:cross_sec} Profile of the dose deposited by the electron beam in the tantalum target calculated for one laser shot.}
\end{figure}

However, isochoric energy deposition would correspond to an extremely short beam duration $\tau_b \ll \tau_s$, where $\tau_s \sim \delta/v_s \approx$ 30 - 90 ps is the characteristic time for stress relaxation in the deposition zone and $v_s$ = 3400 m s$^{-1}$ is the velocity of sound in tantalum. On the contrary, when 
$\tau_b \gg \tau_s$, very slow (isobaric) heating generates thermal dilatation but no pressure rise. In the case of our device, we have $\tau_b \approx$ 50 ns and no significant pressure waves are produced. Moreover, the characteristic time for thermal conduction is $\tau_t \sim \delta^2/D \approx $ 1 - 10 ns, where D $\approx$ 2.5$\times$10$^{-5}$ m$^2$ s$^{-1}$ is the thermal diffusivity of tantalum. Then the interaction of the electron beam with the tantalum anode locally increases the temperature that is completely dissipated by thermal conduction over a  short characteristic time compared to the time between two consecutive laser shots ($\ge$ 100 ms). Therefore, no loss of material from the front of the anode due to melting or hydrodynamic effects is expected.

\subsection{Effects of plasma deposition}

The aluminum plasma produced during the laser-matter interaction has been characterized in previous studies through measurements of charge and energy distributions of the ions\cite{comet2016absolute}. A maximum density of (6$\pm$1) 10$^{14}$ ions sr$^{-1}$ was measured in the symmetry axis of the plasma normal to the target. We showed that the charge densities were not high enough in the aluminum plasma to have significant electron recombination rate in the expanding plasma. This result suggests that the majority of the aluminum atoms emitted from the target are in the ionic state and recombine upon contact with the anode. The anode being placed 5 cm from the target, we deduce that a layer of aluminum of 1 nm thickness is deposited on the anode after about 250 shots. Then, after a few thousands of laser shots the aluminum layer could be thick enough to modify the nuclear excitation yield or to affect the probabilities that conversion electrons exit the anode from the front side. In this section, we propose to study these effects as a function of aluminum deposition thickness.

In a first step, the nuclear $(e,e')$ excitation yields were calculated with the methodology described in section IV. In the code, the anode is composed of an aluminum layer of thickness $x$ deposited on the front side of the tantalum sample 3 $\mu$m thick. The nuclear excitation yields calculated with the DWBA model in an unscreened potential are shown in Fig.11(a) for two target voltages as a function of the aluminum deposition thickness. At 10 kV, where the average electron energy is the lowest, the nuclear excitation yield in the tantalum anode drops by only 10 \% when the deposition thickness increases up to 40 nm. The effect is even less pronounced for a target voltage of 30 kV for which the electron energy loss in the aluminum deposit can be neglected. Therefore, it can be considered that the average nuclear excitation yield remains globally unchanged for each laser shot over a set of at least 10 000 consecutive shots.

\begin{figure}
\begin{minipage}[h]{0.55\linewidth}
\includegraphics[height=7.cm,trim=0 40 300 130, clip=true]{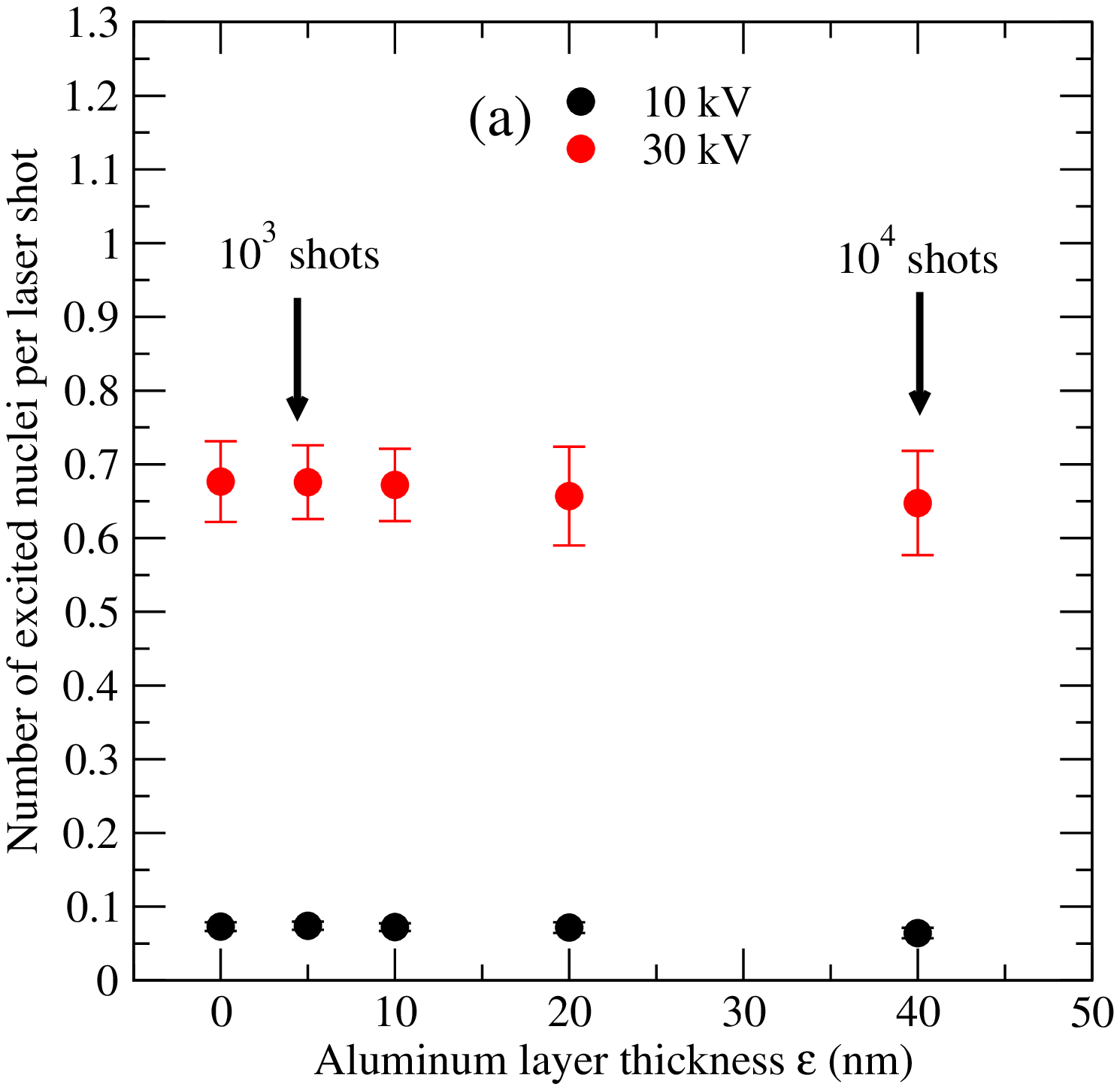}
\end{minipage}
\begin{minipage}[h]{0.4\linewidth}
\includegraphics[height=7.cm,trim=0 40 200 130, clip=true]{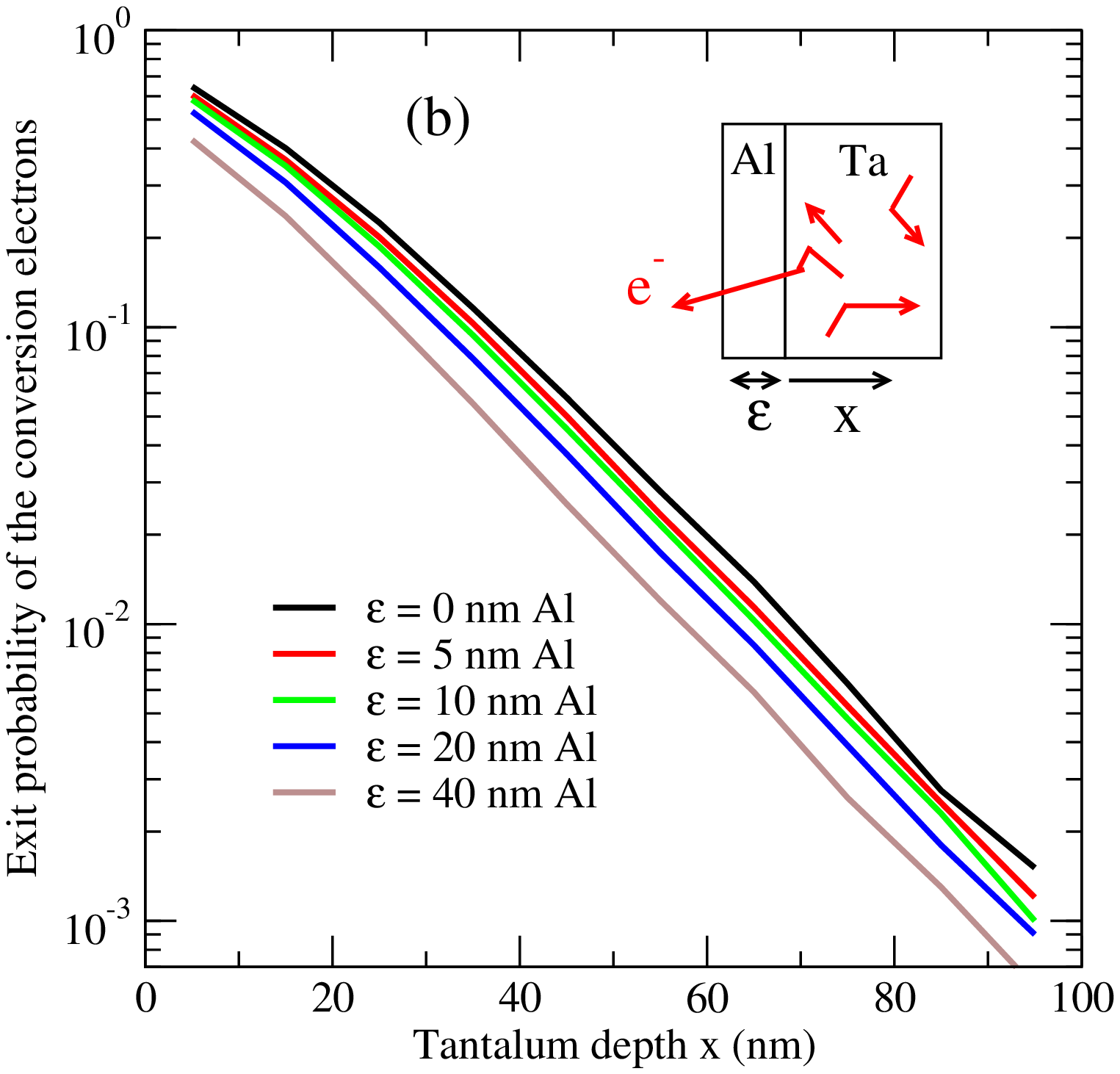}
\end{minipage}
\caption{(a) Number of nuclear excitations in tantalum anode at V$_T$ = 10 and 30 kV as a function of the thickness of the aluminum layer  deposited in front of the anode. Two arrows show the aluminum thicknesses after 10$^3$ and 10$^4$ laser shots. Nuclear excitation yield have been calculated using DWBA (e,e') cross sections of a bare nucleus. (b) Probability of a conversion electron exiting the front side of the tantalum anode as a function of its emission depth and for various aluminum layer thicknesses.}
\end{figure}

In a second step, the probabilities for the conversion electrons to exit the anode front side were recalculated considering the geometry shown in the inset of Fig.11(b). These exit probabilities are plotted in Fig.11(b) as a function of the tantalum depth at which the particle is created for five different aluminum layer thicknesses. As the conversion electrons have a kinetic energy lower than 6 keV, an additional layer of a few nm of aluminum on the front side of the anode can stop the slowest electrons and decrease the exit probability. The fall is even more pronounced the thicker the deposit: the exit probability is twice lower for a 40 nm thick layer, i.e. for a set of about 10 000 laser shots.

For a quantitative experiment, where we would like to limit the variation of the exit probability, we recommend changing the anode, or polishing it to remove the aluminum layer, every 2000 to 3000 laser shots. In that case, the detection probability would fluctuate by less than 20 \%, which remains low compared to the differences of yields of the detectable electrons as calculated in section IV for the different screened (e,e') models.

\section{Conclusion}

In this numerical study, we show that inelastic electron scattering is the main nuclear excitation mechanism occurring in a $^{181}$Ta target irradiated with a new intense 10 - 30 keV electron beam produced using a biased laser-plasma. Through the detection of conversion electrons it could be possible to measure the nuclear excitation yields and thus to constrain the models describing $(e,e')$ cross sections with screened potential. The effects of electron beam heating as well as the plasma deposition on the tantalum target have been quantified, thus allowing the dimensioning of a possible experimental configuration to study $(e,e')$ processes in this range of energy for the first time in neutral atoms. The thickness of the aluminum deposit after a given number of laser shots has to be measured by RBS analysis techniques to precise the number of shots requiring an anode change or treatment.

However, the detection of the $^{181}$Ta decay conversion electrons in an environment of very high background noise presents a strong challenge. The aluminum plasma extends to the anode in a few $\mu$s and it seems unlikely to be able to detect electrons emitted on the first half-life of excited $^{181}$Ta nuclei. Nevertheless a well designed geometry of electrodes submitted to high voltage electrical switches could allow to guide the conversion electrons from the front side of the anode towards an electron detector some 10 $\mu$s after the laser shot. Moreover, these electrons have to be accelerated to energies larger than 10 keV in order to optimize the detection efficiency which can be limited by back-scattering processes.
This efficiency has to be measured in a controlled experimental configuration by implanting in a target, for example, radioactive recoil nuclei decaying via an internal conversion as reported in Ref. \cite{claverie2004search}. 

Moreover, the emission of low-energy electrons following the deposition of a plasma on a catcher foil has already been reported in several works\cite{claverie2004search,bounds1992search}. This exoelectron emission belongs to the group of emission
phenomena which occur during the relaxation of perturbations of the thermodynamic equilibrium in the bulk or at the
surface of a solid.
A counter-experiment performed for example with a tungsten anode will allow to estimate the electron background in the experiment. 
The measurement of time distribution of detected electrons with respect to the time of the laser shot will also be essential in an experiment of this type. The construction of a time spectrum representative of the decay of $^{181}$Ta in its first excited state will sign the nuclear excitation and will allow the extraction of the number of conversion electrons leaving the target or at least determined an upper limit value.

Considering the DWBA model cross sections with a screened potential, about 100 conversion electrons would remain detectable after waiting for a duration of 3 half-lives before the application of an electron guiding electric field. This number, calculated for a set of 10 000 laser shots could be easily increased by considering longer irradiation sequences. At a maximum laser frequency of 10 Hz, 1 million shots can be performed in about 30 hours of acquisition. With this gain of a factor of 100 on the available electrons for detection it would be possible to discriminate the cross sections of one model from another or at least exclude the models giving the most favorable cross sections. 

\begin{acknowledgments}
The authors acknowledge Dr. D. Smith for careful reading of the manuscript.
The research of one of the authors (A.Ya.D.) was supported by a grant of the Russian Science Foundation (Project No 19-72-30014)
\end{acknowledgments}

\medskip

\textbf{DATA AVAILABILITY}

\medskip

The data that support the findings of this study are available from the corresponding author upon reasonable request.

\bibliography{apssamp}% Produces the bibliography via BibTeX.

\end{document}